\documentclass[11pt]{article}
\def\fourth  	{\textstyle {1 \over 4} \displaystyle}
\def\inv16  	{\textstyle {1 \over 16} \displaystyle}
\def\3ov2  	{\textstyle {3 \over 2} \displaystyle}
\def\4ov3  	{\textstyle {4 \over 3} \displaystyle}

\def\beq{\begin{equation}} \def\eeq{\end{equation}}
\def\bea{\begin{eqnarray}} \def\eea{\end{eqnarray}}

       \def\cC{{\cal C}}
\def\cD{{\cal D}}       

\def\Cliff{{\cC\!\ell}}
\def\cF{{\cal F}}
\def\cG{{\cal G}}

\def\cJ{{\cal J}}  	 

\def\cL{{\cal L}} 	\def\bolcL { {\mbox{\boldmath $\cal L$}} }

\def\cN{{\cal N}}

\def\cP{{\cal P}}

\def\cS{{\cal S}}

\def\AG{{\rm Araya-G\'ochez\,}}

 		 	\def\be{ {\bf e}} 	\def\br{ {\bf r}} 	
 	 	 	\def\bc{ {\bf c}} 	
 	 		 	 	
\def\bV{{\bf V}} 	\def\bR{{\bf R}} 	 	 	

\def\TM{{\rm TM}}

\def\bOne { {\mbox{\boldmath $1$}} }
\def\UL{\underline{~}}

 		\def\bvarep { {\mbox{\boldmath $\varepsilon$}} }

\def\Del { {\mbox{\boldmath $\nabla$}\!} }
\def\bdel { {\mbox{\boldmath $\partial$}} }

\begin{document}

\title{On the Vacua of General Relativity in the Backdrop of the Higgs VEV
} 

\author {Rafael A.
Araya-G\'ochez\footnote{\small arayag@tapir.caltech.edu  	{\hfill To Nicole} 
  	}
}

\date{}
\maketitle					

\begin{abstract} 	
The classical notion of manifold may be augmented to institute the algebras associated with the generators of the spinor and vector representations 
of the Lorentz group
on two copies of the cotangent space.  In the former case, this results in the emergence of a Dirac operator out of the vacuum of General Relativity.
In the backdrop of the Higgs VEV, a volume form with such a dual character endows the augmented manifold with dark matter content
as the simplest conceivable geometric structure that could emerge out of the unified vacuum.
\end{abstract}


\begin{quote}
\begin{flushright}
{\footnotesize {\it
Mathematical beauty contradicts physical reality; 	\\	
their perfect marriage is the dream of every theorist.	\\	
To dream is sublime; to wake up, to exist.\\}} 
\end{flushright} 
\end{quote}

\section{Preface}

	The notions discussed in this paper are sufficiently foreign to the physics community that I thought it proper to begin by addressing the standard 
 premises of a modern paper on quantum gravity while framing our work within that construct.
Quantum field theory on flat spacetime contains particles and fields in a unified scheme where the spacetime is fixed and non-dynamical.
 Our notion of ``algebraically augmented manifold"  
 has fermion content but these fields {\it emerge} geometrically as an intrinsic part of the manifold.

	A paper on GR usually starts by postulating that one has a topological, smooth, differentiable manifold.
Our flavor of algebraic manifold is certainly differentiable 
 and it can be made smooth by providing compact support to the emergent spinor content.
 However, given that the local charts of a topological manifold are trivially modeled on Euclidean space, our augmented manifold is definitely {\it not} topological!
 The construction is entirely differential so we will only talk at the level of tangent spaces and Lie algebras.
Our thesis sustains that the tangent space to curved spacetime has a dual character; 
roughly speaking, corresponding to the spinor and vector representation of the Lorentz algebra\cite{AG18a}.
Such a tangent space must be modeled either on Minkowski space \bR$^{1,3}$ or on Clifford space $\Cliff^{1,3}$.

In the language of fiber bundles,
 the world space is the base manifold and when the tangent space is modeled on Clifford space, a Dirac operator emerges out of the vacuum to compensate for 
 Clifford charts that require a partition of unity with appropriately augmented algebraic structure.
 In this case, the world spacetime constitutes a hybrid space: part Clifford, part Minkowski spacetime. 
Geometrically, 
 these statements translate into a 4D volume form with a dual character: 
 One corresponding to the usual vacuum of GR and another with matter content in the form of Majorana bi-spinors\cite{AG18b}.
This is by far our boldest claim.
The core of this paper, \S \ref{sec:EmergDirOps},
 gives credence to the emergence of a Dirac operator as an geometrically intrinsic structure by formally developing calculus on algebraically augmented manifolds.
This is bound to be a controversial issue so the reader in encouraged to trek through that section with due prudence.

\section{Introduction}

	From a purely physical standpoint, we would like to argue that the notion of empty space is vacuous:
 Neither the Lorentz group nor Minkowski space can be defined without inertial observers.
 In this short note, we summarize an important result that gives credence to such a deep philosophical principle.
 The difference between the mathematician's volume form and the physicist interpretation of such a notion is 
 that the latter should understand the {\it tensor density} as the energy density of the vacuum, in some shape or form:
\beq	\int_\cD {\bf \rm Vol}  = \int_\cD \varrho \, d^4x 
\label{Eq:VacuumEnergy}
\eeq
Note that $\varrho$ here denotes the akin to a mass density in Euclidean space but for an ``un-foliated" 4D volume element.
Our aim is to expound the physical meaning of this symbol while maintaining physical balance across the equation.  This is done in \S \ref{sec:ConjOnDE}


	A standard tensor density is a function $\mu: V \times...\times V \rightarrow \bR$ that assigns a volume to the parallelotope 
spanned by n vectors in an n-dimensional vector space.  {\it Modulo orientation,} the tensor density and the volume form may be identified.  
Under a change of coordinates parametrized by a Jacobian matrix $x' = \cJ x$, the tensor density scales as:
\beq	{1 \over {|{\rm det} \cJ|}} \, \mu (\cJ x_1,...,\cJ x_n) = \mu(x_1,...,x_n).
\label{Eq:TensDens}
\eeq
 This construction defines the line bundle of 1-densities where 
${|{\rm det} \cJ|}^{-1}$ belongs to the 1-dimensional representation of the structure group of chart transformations 
 on a standard manifold surrogate to a partition of unity.
Our proposal amounts to exploiting the freedom to choose the Jacobian determinant appropriate to a 1-dimensional presentation of the Lorentz group 
 in either the standard vector representation associated to the vacuum of general relativity {\it or}
 in the spinor representation associated to an exotic Clifford space while maintaining self-consistency in terms of quantum field content.
Furthermore, in the backdrop of the Higgs VEV,
we interpret the interface between these two spaces as a fundamental excitation in the vacuum of the unified theory.

	To clarify the language, let us work out a trivial example.  
 In 3D, the use of spherical coordinates $(r,\theta,\phi)$ may be associated with holonomic coordinates for a spherically symmetric gravitational field.
 In the tangent space, 
 the corresponding local cartesian set $(x,y,z)$ may be likewise associated with a non-holonomic, orthonormal set endowed with the usual Euclidean metric.
The local volume forms are related by:
\[
	dx \, dy \, dz = r^2 \sin\theta \, dr \; d\theta \,d\phi.
\]
The Jacobian matrix 
\[	\cJ = \partial (x,y,z)/\partial(r,\theta,\phi)
\] 
 has determinant $r^2 \sin\theta$ and that's what shows as a density in front of the ``coordinate volume" $dr \, d\theta \,d\phi.$
Note, in particular, that this Jacobian determinant brings length dimensions to the non-dimensional,  coordinate angle {\it differentials}:
 $\theta \rightarrow rd\theta$ and $\phi\rightarrow r\sin\theta \, d\phi$.
 In this trivial case, vierbein facilitate a smooth transition to the orthonormal basis for the tangent space; 
 e.g., $dx = \bvarep^x_\theta d\theta,$ etc., thus bringing length dimensions to the non-coordinate orthonormal basis vectors.

\section{Classical vierbein}

	The tangent space and its dual, the cotangent space, provide the quintessential example of a ``reflexive structure" for a vector space.
 Insofar as this note is concerned, 
 this will simply mean that a complementary pair of elements from TM and T$^*$M span a comprehensive partition of the space with sufficient algebraic structure 
(see below).
 For instance, 
 in Euclidean space the complementary set of basis vectors: $\bdel_i \in \TM$ and co-vectors $dx^i \in \TM^*$ allows us to build a de-composition operator:
\beq
 \bOne_{][} \, \UL = (\bdel_i \otimes dx^i) \,\UL\, ,
\label{Eq:StaRefStr}
\eeq
 which, when applied to an embedded curve in 3D: $\bc$, yields the standard infinitesimal decomposition along a basis of orthogonal cotangent vectors: 
 $ \bOne_{][} \, \bc = dx^i \, \partial_i \bc = d \br (\bc) \equiv \bc_i \, dx^i$. 	
Such an operator, 
 $(\bdel_i \otimes dx^i) \,\UL\,$, comprises a mixed tensor of type (1,1), 	
 e.g., a linear transformation, equivalent to the identity matrix in an appropriate representation of the space
(see, e.g., pg. 248 of Frankel\cite{Frankel04}).
In the presence of a reflexive structure,
 there exists a canonical way to take the inner product between elements of the dual spaces; 
 i.e.,  $dx^i (\partial_j) = \delta^i_j$.  
This fact enables the construction of exterior differential forms which are integrable in the sense that the value of the integral
 over an oriented domain of integration is independent of the parametrization.
Adhering to a philosophy of natural simplicity, one could say that a reflexive structure is more ``primitive" than a Riemann structure. 

	A Riemann space is said to be canonically equipped with a map:
 $g(\TM \oplus \TM) \mapsto \bR$ from two elements of the tangent space to the field of real numbers. 
 Choosing a flat Minkowski metric, 
$\eta = {\rm diag}(-1, \bOne)$, 
endows the space with a causal inner product on elements of the tangent space. 
 This is the mathematician's {\it normed} vector space.  
 To a physicist, it defines the Lorentz invariant proper length for spacetime events.

	 In classical general relativity, the ``world metric": 
\[
g_{ij} = \bvarep_i^a \,\eta_{ab}\, \bvarep^b_j, 
\] is {\it locally} derived from the flat metric 	
 through the introduction of vierbein $\bvarep_i^a$ (a.k.a. frame fields)
 which possess one tangent, globally flat index: ``$a$" and one world, locally curved spacetime index: ``$i$". 
Informally, vierbein may be thought of as the ``square-root of the world metric" whereas the Minkowski metric is attributed to an ``internal" 
 tangent space which is globally flat by definition.  
	In fiber bundle parlance,
 vierbein personify the quintessential vector-valued 1-form: consuming a vector field on the base and bearing a tangent vector as a generic element
 of the fiber in the vector bundle setup.
{\it Exclusive} to the vector representation of the Lorentz group, vierbein may thus be regarded as mediators of a transition from the locally curved space 
 of general relativity to a flat tangent space with Minkowski signature.

\section{From tangent space to Lie algebra: enter {\it the graviform}}

	Evidently, the tangent space is not a Lie algebra; 	
 however, the exterior product of two copies of the cotangent space does take its values in an appropriate representation of a Lie algebra.
In the vector representation of the Lorentz algebra, 
 the proof is a straightforward application of the Grassmann algebra rules for multiplying two copies of 
Minkowski space\footnote{
This little known fact--to the best of my knowledge--is unpublished although John Baez wonderfully funny Quantum Gravity Seminar makes explicit allusion to it. 
(http://math.ucr.edu/home/baez/qg-fall2000/qg2.2.html)
}: 
 Taking the wedge product of two forms that take values in \bR$^{1,3}$ yields a 2-form that takes its values on the Lie algebra of 
SO(1,3).
Indeed, the generators of 4-rotations have a well known matrix representation:
$(\cL^{\mu\nu})_{\alpha\beta} = i (\delta^\mu_\alpha\delta^\nu_\beta - \delta^\mu_\beta \delta^\nu_\alpha)$,
obeying the general form for a Lie algebra of the Lorentz group:
$[\cL^{\mu\nu}, \cL^{\alpha\beta}] = -i ( \eta^{\mu\alpha} \cL^{\nu\beta} - \eta^{\mu\beta} \cL^{\nu\alpha} 
					+ \eta^{\nu\beta} \cL^{\mu\alpha} - \eta^{\nu\alpha} \cL^{\mu\beta} )$.
 Note, in particular, that the existence of a metric is essential to define the Lie algebra as well as the causal space. 

	On the other hand, in the spinor representation of the Lorentz algebra, the generators:
\[	\cS^{ab} = \fourth  [\gamma^a, \gamma^b],
\] make explicit use of upper indexed gamma matrices as basis co-vectors for a ``Clifford space".  
 In this sense, one could say that 
 ``half of the Lie algebra" associated with either the vector or the spinor representations of the Lorentz group can be realized on a tangent space with 
 a dual algebraic character.
 The foundational notion of {\it algebraic manifold}
 poses that in addition to \bR$^{1,3}$, the local charts representing the cotangent space to universal spacetime 
 can also be modeled on Clifford space: $\Cliff^{1,3}$, a 4-dimensional 
 fixed vector space spanned by the upper-indexed gamma matrices as ``Clifford 1-forms".
In \AG 2018a\cite{AG18a} we have argued that from the point of view of mathematical naturalness,
 the dual character to universal space is dictated by a common origin for the Grassmann and Clifford algebras as quotient tensor algebras.	
Algebraically enriched vierbein as 1-graviforms provide a canonical choice of isomorphism between universal space and a tangent space with such a dual character:
 either Minkowski or Clifford.
Furthermore, the wedge product of two such generalized vierbein--a 2-graviform--has its values in an appropriate representation of the Lie algebra.

\section{On the emergence of Dirac operators from the vacua of GR}
 \label{sec:EmergDirOps}

	A deep mathematical fact involves the geometrically intrinsic meaning of an integral 
 as integration of an exterior differential form over a parametrized subset in a standard n-dimensional manifold M$^n$.  
 We follow the arguments leading to the parametrization independence for such an integral as laid out in 
Theodore Frankel's fabulous book. 
 An oriented parametrized point subset is a pair $(U,o; \cF)$ consisting of a oriented region $(U,o)$ in $\bR^n$ and a differential map:
\[	\cF: U \rightarrow {\rm M}^n.
\] 
The integral of a $i$-form over the p-subset is defined invariantly as the pull-back of the $i$-form to the oriented $U$:
\[	\int_{(U,o; \,\cF)} \beta^i = \int_{(U,o)} \cF^* \beta^i.
\]

	As discussed above, in an algebraic manifold the local charts representing the tangent space to universal space 
 can also be modeled on Clifford space $\Cliff^{1,3}$.
Therefore, in addition to oriented regions $(U,o)$ modeled on Euclidean space, 
 we also have oriented charts $(V,o)$ modeled on Clifford space $\Cliff^{1,3}$ and a second differential map:
\[	\cG: V \rightarrow {\rm M}^n.
\] 
 to complement the algebraic structure in the classical notion of manifold.
On the manifold proper, the subset of points mapping to a Clifford chart comprise a distinguish set so we will label them as a q-subset instead.
Furthermore, the oriented parametrized (p,q)-subsets of an algebraically enriched manifold constitute
 a triple $(U,V,o; \cF,\cG)$ made out of oriented regions: 
$(U,o)$ in $\bR^{1,3}$ and $(V,o)$ in $\Cliff^{1,3}$
 and a pair of differential maps: $\cF$ and $\cG$ with dual algebraic character.
Parametrization invariance of the integral of a differential form on a standard manifold M$^n$ 
 makes essential use of the Jacobian matrix associated with a change of coordinates. 	
 If the re-parametrization preserves the orientation of the p-sets, the integral is invariant under re-parametrization of such sets
 (see, 3.1.c of Frankel\cite{Frankel04}).
Thus,
 by postulating that the orientation is preserved under both differential maps $\cF$ and $\cG$, we can expand this result to an algebraically enriched manifold as well.

	Now,
the integral of a $i$-form over the q-subset is defined invariantly as the pull-back of the $i$-form to the oriented $V$:
\[	\int_{(V,o; \,\cG)} \beta^i = \int_{(V,o)} \cG^* \beta^i.
\]
where $\cG^* \beta^i$ lives in Clifford space and the $\cG$-map must compensate for the ``size" of the Clifford chart bellying compact support for the mathematical 
 structure within its confines.

	Let us expound the nature of this structure.

 	Recall that the decomposition of an infinitesimal vector separating two nearby points of an embedded curve along a basis of orthogonal tangent vectors in 3D
 constitutes a mixed tensor product:
 $d\br (\bc) =  \bdel_i \otimes dx^i \, (\bc) = \bc_i \otimes dx^i$,
 where interpretation of $d\br$ as a vector-valued 1-form allows us to write:
 $d\br \,\UL \equiv \bOne_{][} \, \UL$
 in an appropriate representation of the identity matrix involving a reflexive structure.
Let us generalize this sophisticated construction {\it mutatis mutandis} to a 4D, Lorentzian, algebraically enriched manifold as follows.

	Consider a curve embedded in a curved 4D spacetime {\it without slicing} and carry on with the above analysis until the involvement of a reflexive structure.
 We endow the manifold with Clifford structure by assuming that the curve can run into Clifford charts that require 
 a partition of unity with appropriately augmented algebraic structure.
Eq[\ref{Eq:StaRefStr}] is thus supplemented with a suitable reflexive structure for Clifford space:
\[
 \bOne_{][} \, \UL = (\gamma_a \otimes \gamma^a) \otimes (\bdel_i \otimes dx^i) \,\UL.
\]
The meaning of the tensor product in the first parenthesis on the r.h.s. of this equation is the same as that in the second parenthesis but for Clifford space;
i.e., it is like a vector-valued 1-form where
one has effectively picked up a vector field on the base and “projected” it along the fiber direction (within the context of Clifford space, of course).
The tensor product in between the parentheses entwines the two representations into a mixed algebraic structure.
These may be now re-arranged equivariantly across representations: 
\[
 \bOne_{][} \, \UL  = (\gamma^a \otimes \bdel_i ) \,\otimes\,  (dx^i \otimes \gamma_a) \,\UL.
\]
Lastly, in curved spacetime, the interwiner between representations is encapsulated by the vierbein and its inverse:
\beq
 \bOne_{][} \, \UL = (\gamma^a \be_a^i \bdel_i ) \,\otimes\,  ( \gamma_a \bvarep^a_i dx^i ) \,\UL\,,
\label{Eq:IntertStr}
\eeq
 where we have intertwined the two representations into a single reflexive structure for an algebraic augmentation of the space.

	{\it Modulo} a caveat in the orientation of the domains of integration,
the r\^oles of the two new structures on the r.h.s. of Eq [\ref{Eq:IntertStr}] may be interpreted as those of the equivalent structures in Euclidean space
with the following identifications:
\[	\bdel_i \mapsto (\gamma^a \be_a^i \bdel_i ) ~~~{\rm and}~~~ d x^i  	\mapsto ( \gamma_a \bvarep^a_i dx^i ),
\]
 where the {\bf bold} type-setting on the $\bdel$-operator indicates covariant derivative 
hereon\footnote{
\label{FN:ExtCovDer}
The {\bf bold} typesetting on the components of a curve above, $\bc_i$'s, were set to emphasize the fact that these should be regarded as more intrinsic 
than projections onto tangent planes:
 The $\bc_i$'s live on the embedded curve and could be regarded as ``more intrinsic" than the $d x^i$'s who live on a local linear representation of the curve.
In general, detailed scrutiny of the {\it intrinsic} differential structure of a space is made possible by the Lie derivative: $\bolcL{\UL}$, 
 as a generalization of the familiar {\it advective or directional derivative}: $\UL \cdot {\bf d}$.  	
 To probe a space in intrinsic fashion, 
 one should think of this first order gadget as an operator that consumes a vector field while acting on a continuous function to yield a real number. 
 \'Elie Cartan's evolution on this theme--in the fashion prescribed by Charles Ehresmann for so-called internal symmetry--brings
 about the exterior {\it covariant} derivative, $\UL \cdot \Del \,\UL$, 
 where the connection 1-form compensates for the geometry of the  
 {\it synthetically enlarged} bundle space while fundamentally decoupling the ``external" group geometry of the base from that of the fiber. 
In the language of vector-valued p-forms, such ``perpendicular variance" from the linear representation of the tangent space
 is simply embodied in the more intrinsic geometric structure carried by the curve while setting 
$\Del \rightarrow \bdel$.
}.

	Orienting the curve as a 1D (Grassmannian) sub-manifold embedded in the 4D curved spacetime amounts to a choice of direction to decompose the curve on its 
 tangent space.  On a manifold modeled on Euclidean space, the increments on the parametrized p-subset encode the choice of orientation and one should think of 
 the integration process that yields the curve as the inverse of the differential operator embodied by Eq [\ref{Eq:StaRefStr}]: 
A linear functional eating vector fields continuously on the oriented $U$-region modeled on \bR$^n$ to yield real numbers as integration elements: 
\beq 	(\bdel_i \otimes dx^i) \,\UL\,  
	\stackrel{\int}{\longrightarrow}  \; \int \, \UL \; dx^i (\bdel_i) 	
	\mapsto \bR.
\label{Eq:StanMeas}
\eeq
Of course an embedded curve is a linear, integrated structure and we are using the language in the sense of deeming integration and differentiation 
 as inverse procedures; the difference amounts to an affine translation {\it within} either Clifford or Grassmann space. 
Note that the map in Eq[\ref{Eq:StanMeas}]: $\int \, \UL \; dx \mapsto \bR$ defines the standard Riemann measure in Euclidean space.

	Now, pulling back a 1-graviform $\bvarep^a_i dx^i$ to a $V$-region in Clifford space:
\beq 	\cG^* \bvarep^a \mapsto \Cliff^{1,3},
\label{Eq:PBGrav}
\eeq
 is tantamount to the continuous assignment of a linear combination of upper indexed gamma matrices in $\Cliff^{1,3}$;
 tracing a path on Clifford space as an algebraic representation of the tangent space.
 The key assumption of preservation of the orientation in either representation of the tangent space means that we can take this pull back
 and contract it with lower indexed gammas as in the (spatial) covariant structure on the r.h.s. of Eq [\ref{Eq:IntertStr}] to yield real numbers
while preserving the orientation on a distinguished chart:  
\[  	( \gamma_a \bvarep^a_i dx^i ) \stackrel{\int}{\longrightarrow}  \; \int \, \UL\UL \; \bvarep^a_i dx^i (\gamma_a) \mapsto \bR
\]
 The contracted graviform: $\bvarep^a_i dx^i (\gamma_a) \, \UL \mapsto \int \, \UL \; ds,$
 pushed forward from the Clifford chart back onto the manifold partakes on the usual r\^ole of differential increment of independent variable
 within the integral without regard to orientation 
 whereas
 the emergence of a Dirac operator in curved spacetime manifests the nature of the added mathematical structure encapsulated by the $\cG$-map:
\beq	(\gamma^a \be_a^i \bdel_i ) \,\otimes\,  ( \gamma_a \bvarep^a_i dx^i ) \, \UL\UL 
	\rightarrow
	\int  \gamma^a \be_a^i \bdel_i  \, ds \mapsto \bR, 
\label{Eq:AG-Meas}
\eeq 
where the arrow indicates inverse process and ``$s$" 
becomes an affine parameter along the worldline of a massive, electrically neutral new fermion: a dark matter particle! 
Note that the map in Eq[\ref{Eq:AG-Meas}] above: 
	$\int  \gamma^a \be_a^i \bdel_i  \, ds \mapsto \bR,$ 
defines the equivalent to the Riemann measure for Clifford space. 

\section{A geometrically intrinsic Lagrangian density for dark matter}
 \label{sec:GIL}

	To recapitulate, in an algebraic Lorentzian manifold the integration of an embedded 1D ``sub-manifold"
makes intrinsic use of oriented parametrized subsets in Clifford space representing the tangent space at distinguished charts of the spacetime.
Orienting an arbitrary curve in the 4D curved spacetime and exploring its geometric structure as the curve runs into a Clifford chart
 reveals the emergence of a Dirac operator as a geometrically intrinsic constituent of the spacetime.  
Once such an operator appears, we must interpret the oriented curve as the worldline of a massive fermion;
i.e., the Dirac operator acts as the seed for a Lagrangian density 1-form
 effectively justifying mass insertions\cite{DHM08} in the propagator of the associated fermion (see below).

	In the above sense,
 the appearance of the Dirac structure implies a local slicing of the spacetime along space-like hypersurfaces orthogonal to the local worldline 
of the emerging spinor.
 This is the geometrical meaning of ``on-mass--shell physics" for fermions.
In the simplest possible scenario,
 the time-dependent vector field corresponding to the worldline of a 
4D Majorana bispinor\cite{AG18b} 
 does not generate a one parameter group of diffeomorphisms of the base spacetime as a manifold: It does not satisfy the flow properties
{\bf i-} $\phi_t(\phi_s(p)) = \phi_{t+s} (p)= \phi_s(\phi_t(p))$ and 
{\bf ii-} $\phi_{-t}(\phi_t(p)) = p$.  	
 This drives at the core of the notion of coordinate independence:  
 A time-dependent vector field is a flow in the product manifold $\bR \times M^3$ {\it instead}, where $t$ is the coordinate for $\bR$ (see \S {\bf 4.3b} of Frankel).
 The selection of such a preferred local coordinate breaks the parametrization independence of the 4D spacetime in accordance with the interpretation of 
 the Lagrangian 3-density: $L$, as a 1-form: $L\,dt$ in the 4D curved spacetime (\S {\bf 16.4b} of Frankel).

	From Eq [\ref{Eq:AG-Meas}] above, we read that the change of measure induced by a {\it naked singular geometric point} is operationally equivalent to:
\beq	\int_\cD \varrho \, d^4x 
	\rightarrow 	
	\int_\cD \bar{\varrho}^{^{1/2}} [\gamma \otimes \bdel] \, \varrho^{^{1/2}} d^4x ,
\label{Eq:Majoranas}
\eeq 		
 where the introduction of distributional 
half-densities\footnote{
In the context of geometric quantization,
these spinors may be graciously regarded as Lorentz covariant ``half-densities" on a {\it discrete} metaplectic manifold;
see, e.g., Guillemin \& Sternberg {\it ``Geometric Asymptotics"} (1977).  
At present, it is unclear whether a Lorentz covariant variance of the latter may be naturally identified with the Clifford space.
}
as a conjugate pair of Majorana spinors  	
parametrize the non-locality of the formalism; i.e., roughly speaking, bellying the size of the Clifford chart.  
 The meaning of the tensor product here deserves expounding.  
 Motivation for the introduction of inverse vierbein as intertwiners across the spinor and vector representations of the Lorentz group 
 essentially involves interpretation of the entwined structure:  
 $[\gamma \otimes \bdel] \equiv \gamma^a \be_a^i \bdel_i$ 
 as a contracted reflexive structure in a hybrid space: part Clifford, part Minkowski spacetime.
 This structure is formally dual to the contracted graviform in Eq [\ref{Eq:AG-Meas}].
Furthermore,  
 the implied contraction across representations effectively swaps the volume scale in favor of a mass scale associated with the appearance of the Dirac structure. 
 It is for this reason that we label these as ``points" as opposed to charts.
The points are singular because of they puncture holes in the standard topology of \bR$^{1,3}$.
 Lastly, they are naked because
in the backdrop of the Higgs VEV, 
 the insertion of a Dirac operator within the tensor density in Eq [\ref{Eq:VacuumEnergy}]
 not only ruptures the former into geometric “half-densities” but also generates a mass term recursively in the physical fermion propagator
 effectively dressing it with self-interaction vertices as mass insertions 
 upon refinement of the Feynman rules to two-component, chiral Weyl-van der Waerden 
spinors\cite{DHM08}.  
We interpret this as a fundamental excitation in the vacuum of the unified theory.


\section{And a conjecture on dark energy}
 \label{sec:ConjOnDE}

	With the freedom to change measure to and from Clifford space at a singular geometric point, 
 we can write a general, {\it invariant} volume form for universal Lorentz space as a 4-graviform:
\beq 	{\bf Vol} = \epsilon_{abcd} \, \bvarep^a \wedge \bvarep^b \wedge \bvarep^c \wedge \bvarep^d,
\label{Eq:UnvVolForm}
\eeq
 where each algebraically enriched vierbein, $\bvarep^a$, is a choice of isomorphism between universal space and the cotangent space. 	
Evidently, the breaking of affine symmetry endows spacetime with inhomogeneous structure. 
 We have stressed on the operational nature of these manipulations because the introduction of distinguished charts in the spacetime clearly violates 
 foundational premises in the definition of classical manifolds and tensor densities.
 Concretely, we need to give up translational invariance and the notion of points on the manifold.
Recall that the latter are formally defined as being invariant under a change of coordinate chart surrogate to a partition of unity on the manifold.

	Now, removal of the Dirac operator in Eq [\ref{Eq:AG-Meas}] leaves a conjugate pair of Majorana spinor wavefunctions:
\[	
	\int_\cD \bar{\varrho}^{^{1/2}} [\gamma \otimes \bdel] \, \varrho^{^{1/2}} d^4x 
	\rightarrow \int \bar{\varrho}^{^{1/2}}  \varrho^{^{1/2}} d^4x,
\]
 a scalar 4-density in the spinor representation of SL(2,C) that--modulo a
caveat\footnote{
Note that once the Dirac operator is removed, there exists no {\it a priori} local slicing of the spacetime.  
Dark energy involves interpretation of the spatial 3-density as the minimum of the potential.  
In the absence of a global foliation, the scalar 4-density lacks a clear physical interpretation. 
Such global issues lie beyond the scope of the present paper.
} on the global foliation of the space--may be interpreted
 as a virtual fermion condensate: a {\it Majorana Sea} justifying the vacuum state of the Universe.
 This is the meaning we attach to the tensor density displayed in Eq [\ref{Eq:VacuumEnergy}] of the introduction.

	It seems rather revealing that the density of volume encapsulated by an abstract volume form could embody the full physical content of
 non-compact, causal, and spin spacetime.  Furthermore, exclusive recourse to the Lorentz group and its double cover
 as the structure group of universal space results in a formulation that is manifestly Lorentz covariant throughout. 


{\footnotesize
\begin{flushright} {\it --$\cP\!$ro $\cG$loria $\cN\!$umen $\!${\cal \AE}theris--} \end{flushright}
} 

\vfill
{\footnotesize
\noindent{\bf Acknowledgements:} \\
{\noindent
I am particularly grateful to John Schwarz for intellectual exchange throughout this work which led to the initial conception 
and evolution of the key notion of graviforms.  Roman Buniy and Tudor Dimofte provided key criticism of prior manuscripts.
I also thank my teachers: Anton Kapustin, Hirosi Ooguri, Anton Geraschenko and 
Guy de Teramond from whom I benefited greatly in this marvelous journey.  Furthermore, this work would not have been possible 
without the intellectual legacy of two great minds in Mathematics: Shlomo Sternberg and Michael Artin.  To them, I extend a great many thanks. 
Very helpful comments on preliminary manuscripts were given by Jeff Rabin.
} }


\end{document}